\documentclass[conference]{IEEEtran}
\IEEEoverridecommandlockouts
\usepackage{amsmath, amssymb, amsfonts, bm, mathtools}
\usepackage{amsthm}

\usepackage[version=4]{mhchem}
\usepackage{chemfig}

\usepackage{algorithm}
\usepackage{algorithmic}

\usepackage{bbm}
\usepackage{graphicx}
\usepackage{subfigure}
\usepackage{svg}
\usepackage{tikz}
\usepackage[american]{circuitikz}
\usetikzlibrary{
  arrows.meta, 
  decorations.markings, 
  circuits, 
  automata,
  arrows, 
  positioning, 
  calc
}

\usepackage{booktabs}
\usepackage{makecell}

\usepackage{textcomp}
\usepackage{soul}       
\usepackage{verbatim}
\usepackage{xcolor} 

\usepackage{cite}
\usepackage[hidelinks,bookmarks=false]{hyperref}
\usepackage[capitalise,noabbrev]{cleveref}
\Crefformat{figure}{#2Fig.~#1#3}
\Crefformat{equation}{#2Eq.~(#1)#3}
\crefname{equation}{Eq.}{Eqs.}
\crefname{figure}{Fig.}{Figs.}

\usepackage{xr}
\usepackage{xr-hyper} 

\usepackage[
  range-phrase=--,
  per-mode=symbol-or-fraction,
  range-units=single,
  list-units=single,
  detect-all,
  list-final-separator={, and },
]{siunitx}

\usepackage{acro}
\usepackage{mfirstuc}

\DeclareAcronym{MC}{short=MC, long=molecular communication}
\DeclareAcronym{MCvD}{short=MCvD, long=molecular communication via diffusion}
\DeclareAcronym{IM}{short=IM, long=information molecule, short-plural=s, long-plural=s}
\DeclareAcronym{TX}{short=TX, long=transmitter, short-plural=s, long-plural=s}
\DeclareAcronym{RX}{short=RX, long=receiver, short-plural=s, long-plural=s}
\DeclareAcronym{IoBNT}{short=IoBNT, long=Internet of Bio-Nano Things}
\DeclareAcronym{ssDNA}{short=ssDNA, long=single-stranded DNA}
\DeclareAcronym{cDNA}{short=cDNA, long=complementary DNA}

\DeclareAcronym{EM}{short=EM, long=electromagnetic}
\DeclareAcronym{FSO}{short=FSO, long=free-space optical}

\DeclareAcronym{ML}{short=ML, long=maximum-likelihood}
\DeclareAcronym{LS}{short=LS, long=least-squares}
\DeclareAcronym{NNLS}{short=NNLS, long=non-negative least squares}
\DeclareAcronym{MAP}{short=MAP, long=maximum a posteriori}
\DeclareAcronym{MMSE}{short=MMSE, long=minimum mean squared error}
\DeclareAcronym{MIMO}{short=MIMO, long=multiple-input multiple-output}
\DeclareAcronym{CDMA}{short=CDMA, long=code division multiple access}

\DeclareAcronym{MSE}{short=MSE, long=mean squared error}
\DeclareAcronym{FIM}{short=FIM, long=Fisher information matrix}
\DeclareAcronym{CRB}{short=CRB, long=Cramér-Rao bound}

\DeclareAcronym{SNR}{short=SNR, long=signal-to-noise ratio}
\DeclareAcronym{SINR}{short=SINR, long=signal-to-interference-plus-noise ratio}

\usepackage{standalone}

\usepackage{listofitems} 
\usepackage[outline]{contour} 
\contourlength{1.4pt}

\colorlet{myred}{red!80!black}
\colorlet{myblue}{blue!80!black}
\colorlet{mygreen}{green!60!black}
\colorlet{myorange}{orange!70!red!60!black}
\colorlet{mydarkred}{red!30!black}
\colorlet{mydarkblue}{blue!40!black}
\colorlet{mydarkgreen}{green!30!black}

\usepackage{titlesec}
\titlespacing\section{0pt}{6pt plus 2pt minus 2pt}{4pt plus 2pt minus 2pt}
\titlespacing\subsection{0pt}{4pt plus 2pt minus 2pt}{3pt plus 2pt minus 2pt}

\begin{document}
\bstctlcite{IEEEexample:BSTcontrol}
\title{Molecular ISAC via Markov State--Space Modeling: Joint Distance Sensing and Data Detection}

\author{
    \IEEEauthorblockN{
        Ruifeng Zheng\IEEEauthorrefmark{1},
        Pengjie Zhou\IEEEauthorrefmark{1},
        Martín Schottlender\IEEEauthorrefmark{1},
        Veronika Volkova\IEEEauthorrefmark{1},
        Juan A. Cabrera\IEEEauthorrefmark{1}\IEEEauthorrefmark{2}, \\
        Frank H.\,P. Fitzek\IEEEauthorrefmark{1}\IEEEauthorrefmark{2}, and 
        Pit Hofmann\IEEEauthorrefmark{1}\IEEEauthorrefmark{2}
    }
    \IEEEauthorblockA{
         \IEEEauthorrefmark{1}Deutsche Telekom Chair of Communication Networks, Dresden University of Technology, Germany\\
         \IEEEauthorrefmark{2}Centre for Tactile Internet with Human-in-the-Loop (CeTI), Dresden, Germany\\
         \texttt{\{firstname.lastname\}@tu-dresden.de}
     }
    \thanks{This work was funded by the German Research Foundation (DFG, Deutsche Forschungsgemeinschaft) as part of Germany’s Excellence Strategy – EXC 2050/2 – Project ID 390696704 – Cluster of Excellence “Centre for Tactile Internet with Human-in-the-Loop” (CeTI) of Dresden University of Technology.
    The authors also acknowledge the financial support by the Federal Ministry of Research, Technology and Space (BMFTR) of Germany in the program “Souverän. Digital. Vernetzt.” Joint project 6G-life, project identification number 16KIS2413K, the program “Verbundprojekt: Disruptive Kommunikationsparadigmen für technologische Souveränität, Resilienz und Shared Prosperity - Translation in Industrie und Aufbau innovativer Technologiedemonstratoren - CommUnity,” project identification number 16KISS012K, and the project IoBNT, grant number 16KIS1994.}
}

\maketitle

\begin{abstract}
This paper develops a molecular integrated sensing and communication (ISAC) framework that exploits the same molecular observations for physical-parameter sensing and data detection. As a representative instantiation, we consider a microfluidic molecular communication (MC) channel and study transmitter--receiver (TX--RX) distance sensing, where the distance affects the propagation delay, transient response, and inter-symbol interference structure. A distance-parameterized Markov state--space model is established to obtain distance-dependent channel impulse responses and a block observation model for on-off keying signaling. Based on this model, we design a pilot-assisted low-complexity receiver that combines distance initialization, decision-feedback equalization (DFE), and iterative joint refinement. Numerical results show accurate distance sensing and improved bit error ratio (BER), demonstrating the mutual benefit between sensing and communication and highlighting microfluidic MC as a representative platform for molecular ISAC.
\end{abstract}

\begin{IEEEkeywords}
Integrated sensing and communication (ISAC), molecular communication (MC), state-space model, distance sensing, data detection.
\end{IEEEkeywords}

\section{Introduction} \label{sec:intro}

\IEEEPARstart{M}{olecular} communication (MC) has emerged as a promising communication paradigm for microscale and nanoscale systems operating in environments where conventional electromagnetic signaling is ineffective or impractical~\cite{farsad2016comprehensive,lu2020wireless,akan2023internet}. By using molecules as information carriers, MC is inherently compatible with biochemical environments and has attracted growing interest in applications such as biosensing, in-body communication, and molecular information processing~\cite{gomez2024dna,zheng2025DNA-Based}.

A fundamental feature of MC is that the received molecular signal is jointly determined by the transmitted symbols and the underlying physical channel parameters governing molecular transport and reception. Such parameters may include the transmitter--receiver (TX--RX) distance, flow conditions, reaction rates, or environmental properties. Therefore, the same molecular observations can, in principle, be exploited not only for data detection but also for sensing or inferring physical channel characteristics. This intrinsic dual use of the molecular signal suggests that MC naturally supports a general integrated sensing and communication (ISAC) perspective.

While ISAC has become an important paradigm in wireless communications, where shared waveforms and hardware resources are jointly exploited for sensing and communication~\cite{liu2020joint,zhang2021overview,lu2024integrated}, its formulation in the MC domain remains largely unexplored. Existing MC studies mainly focus on channel modeling, modulation, detection, and estimation under fixed or known channel conditions~\cite{yilmaz2014three,zheng2025anis,zheng2026molecular}. Although related sensing tasks, such as distance estimation, have been investigated in specific MC settings~\cite{wang2015distance}, a unified molecular ISAC framework that explicitly couples channel-parameter sensing and symbol detection within a common observation model remains lacking.

In this paper, we take a first step toward such a molecular ISAC framework by considering MC systems in which the received molecular response depends jointly on the transmitted symbols and physical channel parameters. Under this view, channel-parameter sensing and data detection can be performed using the same molecular observations. As a representative and analytically tractable instantiation, we consider a microfluidic MC channel and focus on TX--RX distance sensing, where the distance directly affects the propagation delay, transient response, and inter-symbol interference (ISI) structure. 

Based on this setting, we develop a distance-parameterized Markov state--space model that maps each candidate TX--RX distance to a corresponding molecular channel response. The resulting formulation yields distance-dependent channel impulse responses (CIRs) and a discrete-time block observation model that preserves the transient response within each symbol interval. Building on this common observation model, we design a pilot-assisted low-complexity receiver that performs distance initialization, distance-aware decision-feedback detection, and iterative joint refinement. This design is relevant to IoBNT-inspired lab-on-chip and microfluidic biosensing systems, where molecular observations may support data recovery, TX--RX geometry calibration, proximity awareness, and adaptive detection.

The main contributions of this paper are summarized as follows. 
\textit{(1)} We introduce a molecular ISAC perspective that exploits the same molecular observations for both channel-parameter sensing and data detection. 
\textit{(2)} We develop a distance-parameterized Markov state--space model and a corresponding block observation model for microfluidic MC. 
\textit{(3)} We design a pilot-assisted distance-aware receiver with decision-feedback equalization (DFE) and iterative joint refinement, and demonstrate through numerical results that sensing and communication can mutually improve each other.

\section{Distance-Parameterized Markov Channel Model}
\label{sec:model}

This section develops the distance-parameterized Markov channel model that provides the physical basis for the proposed molecular ISAC framework. The key idea is to represent the molecular channel response as a function of the TX--RX distance, so that the same observation model can be used for both distance sensing and data detection. To instantiate this idea, we consider a representative microfluidic MC channel and embed the TX--RX distance into the Markov transition structure through the receiver location.

\begin{figure}
    \centering
    \includegraphics[width=0.8\linewidth]{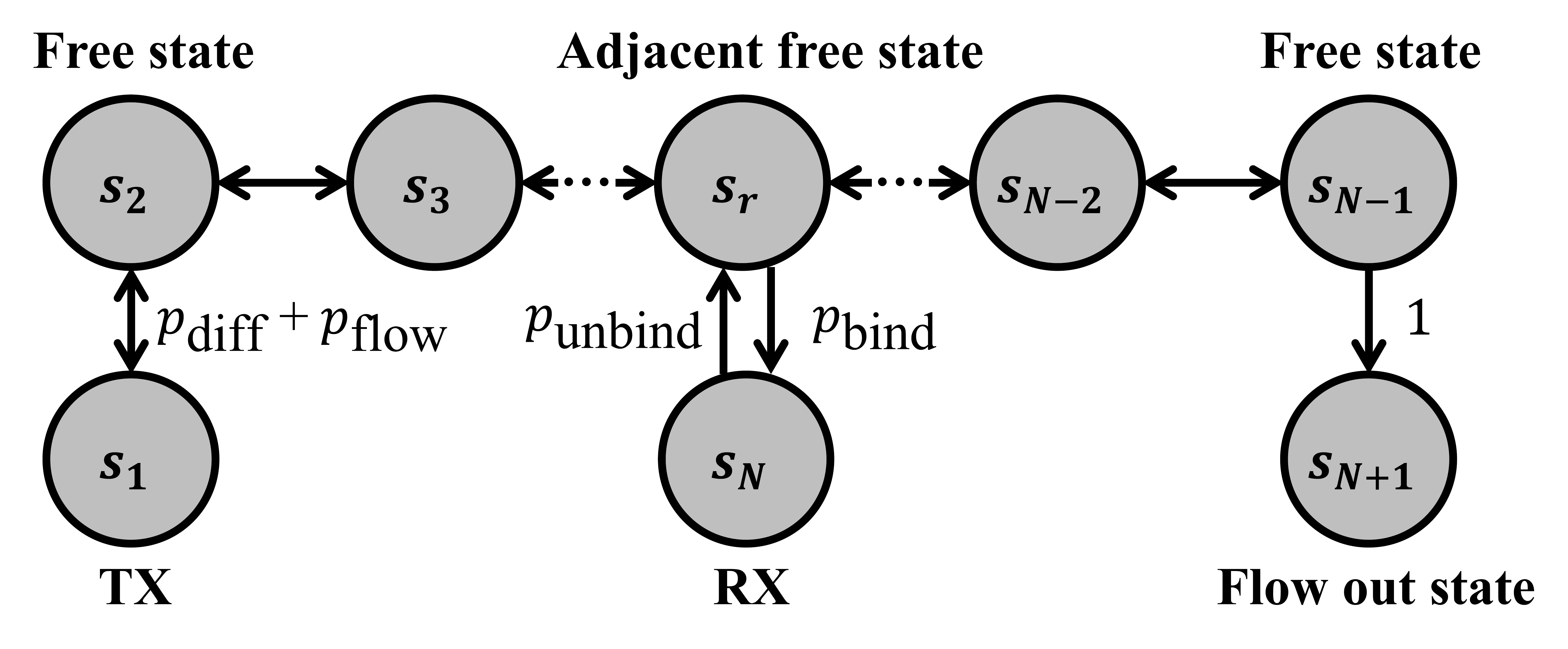}
    \caption{Markov chain representation of the microfluidic MC channel.}
    \label{fig:markov_chain}
\end{figure}

\subsection{Distance-Dependent Markov Transition Structure}
\label{subsec:microfluidic_markov_model}

Building on the microfluidic MC model in~\cite{zheng2026system}, we consider a microfluidic channel in which information molecules propagate along the flow direction under diffusion, downstream advection, reversible binding at the receiver surface, and irreversible flow-out loss. The channel is represented by a discrete-time Markov chain. Although this paper focuses on a one-dimensional implementation for clarity and tractability, the same principle can be extended to higher-dimensional Markov grids by increasing the state dimension.
The state space is defined as
\begin{equation}
    \mathcal{S}=\{s_1,\ldots,s_N,s_{N+1}\}.
    \label{eq:state_space}
\end{equation}
The states $s_1,\ldots,s_{N-1}$ denote free propagation states along the channel, $s_N$ denotes the receiver-associated bound state, and $s_{N+1}$ denotes the absorbing flow-out state; see~\cref{fig:markov_chain}. Molecules are released into the inlet state $s_1$. The receiver is represented by the free--bound state pair $(s_r,s_N)$, where $s_r$ is the free state adjacent to the receiver surface and $s_N$ is the corresponding bound state.

Taking the inlet state $s_1$ as the spatial origin, the TX--RX distance is parameterized by the receiver index $r$ as
\begin{equation}
    d=(r-1)\Delta x,
    \label{eq:distance_receiver_index}
\end{equation}
where $\Delta x$ is the spatial step size. Thus, changing the TX--RX distance is equivalent to changing the location of the receiver-adjacent free state $s_r$. The elementary transport and reaction probabilities remain fixed for given physical parameters, while the receiver-associated binding and unbinding transitions are placed at different positions in the Markov chain. As a result, each candidate distance induces a distinct distance-dependent transition matrix.

For a given candidate distance $d$, the corresponding one-step transition matrix is defined as
\begin{equation}
    \mathbf{P}(d)\triangleq [p(i,j)] \in \mathbb{R}^{(N+1)\times(N+1)}, 
    \label{eq:P_def_model}
\end{equation}
where $p(i,j)$ denotes the probability that a molecule transitions from state $s_j$ to state $s_i$ within one Markov sampling interval $\Delta t$. Since probability is conserved, $\mathbf P(d)$ is column-stochastic. Separating the transient states from the absorbing flow-out state, $\mathbf P(d)$ can be written as
\begin{equation}
    \mathbf{P}(d)=
    \begin{bmatrix}
        \mathbf{Q}(d) & \mathbf{0}_{N\times 1}\\
        \bm{\psi}^{\top}(d) & 1
    \end{bmatrix},
    \label{eq:P_block_model}
\end{equation}
where $\mathbf{Q}(d)\in\mathbb{R}^{N\times N}$ describes the transitions among the transient states and $\bm{\psi}(d)\in\mathbb{R}^{N\times 1}$ collects the transition probabilities from the transient states to the absorbing flow-out state.

The nonzero entries of $\mathbf Q(d)$ and $\bm\psi(d)$ are determined by four elementary transition probabilities. The diffusion and downstream-advection probabilities are given by
\begin{equation}
    p_{\mathrm{diff}}=\frac{D\Delta t}{(\Delta x)^2},
    \qquad
    p_{\mathrm{flow}}=\frac{v\Delta t}{\Delta x},
    \label{eq:pdiff_pflow_model}
\end{equation}
where $D$ is the diffusion coefficient and $v$ is the mean flow velocity. The reversible receiver reaction is described by
\begin{equation}
    p_{\mathrm{bind}}=k_{\mathrm{on}}c_{\mathrm{p}}\Delta t,
    \qquad
    p_{\mathrm{unbind}}=k_{\mathrm{off}}\Delta t,
    \label{eq:pbind_punbind_model}
\end{equation}
where $k_{\mathrm{on}}$, $k_{\mathrm{off}}$, and $c_{\mathrm p}$ denote the association rate, dissociation rate, and receptor concentration, respectively. The discretization is chosen such that the resulting transition probabilities are nonnegative and the total outgoing probability from each state does not exceed one.

\subsection{State-Space Channel Model}
\label{subsec:state_space_model}

Given the distance-dependent transient-state transition matrix $\mathbf{Q}(d)$, the microfluidic channel is described by a linear state-space model. The vector $\mathbf{x}_k(d)\in\mathbb{R}^{N}$ represents the expected numbers of molecules in the transient states at Markov sampling step $k$, and evolves as
\begin{equation}
    \mathbf{x}_k(d)
    =
    \mathbf{Q}(d)\mathbf{x}_{k-1}(d)
    +
    \mathbf{b}\,u_{k-1},
    \label{eq:state_model_final}
\end{equation}
where $u_k$ is the molecular input and
$\mathbf{b}=[1,0,\ldots,0]^{\top}$ is the input vector, since molecules are injected into the inlet state $s_1$. The receiver output is the expected number of bound molecules,
\begin{equation}
    y_k(d)=\mathbf{h}^{\top}\mathbf{x}_k(d),
    \label{eq:obs_model_final}
\end{equation}
where $\mathbf{h}=[0,0,\ldots,1]^{\top}$ selects the receiver-associated bound state. The absorbing flow-out state is not directly observed but accounts for irreversible molecular loss and ensures probability conservation.

\begin{figure}[t]
    \centering
    \includegraphics[width=0.8\linewidth]{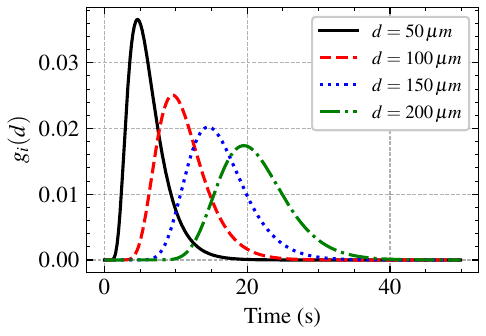}
    \caption{Distance-dependent Markov-step CIRs $g_i(d)$ for different TX--RX distances.} 
    \label{fig:cir_distance}
\end{figure}

For an initially empty channel with $\mathbf{x}_0=\mathbf{0}$, repeated substitution of~\cref{eq:state_model_final} gives
\begin{equation}
    \mathbf{x}_k(d)
    =
    \sum_{i=0}^{k-1}
    \mathbf{Q}(d)^i \mathbf{b}\,u_{k-i-1}.
    \label{eq:iterative_state_model_final}
\end{equation}
Therefore, the distance dependence of $\mathbf Q(d)$ directly induces a distance-dependent input--state relation. For the considered single-input single-output (SISO) setting, the Markov-step CIR is defined as
\begin{equation}
    g_i(d)\triangleq \mathbf{h}^{\top}\mathbf{Q}(d)^i\mathbf{b},
    \qquad i\in\mathbb{Z}_{\ge 0}.
    \label{eq:cir_def_model}
\end{equation}
The quantity $g_i(d)$ represents the expected receiver response at step $i$ due to a unit molecular release. Since the receiver location changes the transition matrix $\mathbf Q(d)$, different TX--RX distances produce different CIRs, as illustrated in~\cref{fig:cir_distance}. These distance-dependent CIRs form the basis for the observation templates used for joint distance sensing and data detection.

\subsection{Discrete-Time Observation and Symbol-Wise Processing}
\label{subsec:discrete_time_obs_model}

We next construct the discrete-time observation model induced by the distance-dependent CIR in~\cref{eq:cir_def_model}. The objective is to map the transmitted OOK symbol sequence to noisy receiver observations and then organize the retained samples into symbol-wise blocks. These blocks preserve transient features that are informative for both distance sensing and data detection. 
For OOK signaling, each symbol interval $T_b$ contains $N_s \triangleq T_b/\Delta t \in \mathbb{Z}_{>0}$ Markov sampling steps. The transmitted symbols satisfy $a_m\in\{0,1\}$ for $m=1,2,\ldots, K$. 
Under symbol-synchronous impulsive release, bit-$1$ releases a number of $N_{\mathrm{tx}}$ molecules at the beginning of the corresponding symbol interval, whereas bit-$0$ releases no molecules. Hence, the molecular input sequence is
\begin{equation}
    u_k=\sum_{m=1}^{K} N_{\mathrm{tx}} a_m\,\delta[k-(m-1)N_s],
    \label{eq:ook_step_input}
\end{equation}
where $\delta[\cdot]$ denotes the discrete-time Kronecker delta function. 

Using the distance-dependent CIR in~\cref{eq:cir_def_model}, the noiseless receiver response under a candidate distance $d$ is
\begin{equation}
    \mu_k(d,\mathbf a)
    =
    \sum_{m=1}^{\left\lfloor (k-1)/N_s \right\rfloor+1}
    N_{\mathrm{tx}} a_m\,
    g_{k-(m-1)N_s-1}(d),
    \label{eq:step_mean_obs_model}
\end{equation}
where $\mathbf a=[a_1,a_2,\ldots,a_K]^\top$. The corresponding receiver observation is modeled as
\begin{equation}
    z_k=\mu_k(d,\mathbf a)+w_k.
    \label{eq:step_obs_model}
\end{equation}
For analytical tractability, the noise term is approximated as Gaussian,
\begin{equation}
    w_k \sim \mathcal{N}(0,\sigma_k^2),
    \label{eq:step_gaussian_noise_model}
\end{equation}
with signal-dependent variance~\cite{hassibi2005biological,zheng2026markov}
\begin{equation}
\begin{aligned}
    \sigma_k^2
    &\approx
    \sum_{m=1}^{\left\lfloor (k-1)/N_s \right\rfloor+1}
    N_{\mathrm{tx}} a_m\,
    g_{k-(m-1)N_s-1}(d) \\
    &\quad \times
    \bigl(1-g_{k-(m-1)N_s-1}(d)\bigr).
\end{aligned}
\label{eq:step_variance_model}
\end{equation}
This approximation captures the signal-dependent counting fluctuation of the received molecules while keeping the observation model tractable. In the considered microfluidic MC system, downstream flow and flow-out loss shorten the molecular residence time, thereby reducing long-range temporal dependence among retained samples. Hence, the receiver uses marginal noise statistics, while covariance-aware block processing is left for future work.

To preserve the distance-dependent transient response within each symbol interval, we sample the receiver output multiple times per symbol. Specifically, $T_b$ denotes the symbol interval, $T_s$ denotes the retained sampling interval, and $X$ denotes the number of retained samples within one symbol interval. We choose
\begin{equation}
    T_s \triangleq M\Delta t, \quad T_b=XT_s,
    \qquad M,X\in\mathbb{Z}_{>0},
    \label{eq:Ts_X_def}
\end{equation}
so that $X=N_s/M$ samples are retained per symbol. For the $m$-th symbol interval, the retained Markov sampling indices are
\begin{equation}
    k_{m,q}\triangleq (m-1)N_s+qM,
    \qquad q=1,\ldots,X.
    \label{eq:block_sample_index}
\end{equation}
The corresponding observation block is modeled as
\begin{equation}
    \mathbf z_m
    =
    \boldsymbol{\mu}_m(d,\mathbf a)+\mathbf w_m
    \in \mathbb{R}^{X},
    \label{eq:block_obs_model}
\end{equation}
where $\mathbf z_m=[z_{k_{m,1}},\ldots,z_{k_{m,X}}]^\top$ is the received block, $\boldsymbol{\mu}_m(d,\mathbf a)=[\mu_{k_{m,1}}(d,\mathbf a),\ldots,\mu_{k_{m,X}}(d,\mathbf a)]^\top$ is the noiseless block mean, and $\mathbf w_m=[w_{k_{m,1}},\ldots,w_{k_{m,X}}]^\top$ is the block noise vector.

\section{Joint Distance Sensing and Data Detection}
\label{sec:isac}

This section develops a low-complexity receiver for joint distance sensing and data detection. The receiver exploits the coupling between the TX--RX distance and the transmitted symbol sequence in the received molecular observations. Specifically, distance sensing provides channel templates for data detection, while detected data symbols provide additional observations for distance refinement. To avoid the prohibitive complexity of an exhaustive joint search over the distance and data symbols, we adopt an alternating design consisting of pilot-assisted distance initialization, distance-aware decision-feedback detection, and iterative joint refinement.

\subsection{Pilot-Assisted Distance Initialization}
\label{subsec:pilot_init}

The transmission frame contains $K=K_{\mathrm p}+K_{\mathrm d}$ symbols, including the known pilot sequence $\mathbf a^{\mathrm p}=[a_1^{\mathrm p},\ldots,a_{K_{\mathrm p}}^{\mathrm p}]^\top$ and the unknown data sequence $\mathbf a^{\mathrm d}=[a_1^{\mathrm d},\ldots,a_{K_{\mathrm d}}^{\mathrm d}]^\top$. Thus, the complete transmitted symbol sequence is
\begin{equation}
    \mathbf a
    =
    [(\mathbf a^{\mathrm p})^\top,(\mathbf a^{\mathrm d})^\top]^\top .
    \label{eq:frame_symbol_sequence}
\end{equation}
The candidate TX--RX distance set is denoted by $\mathcal D$. Since the pilot symbols $\mathbf a^{\mathrm p}$ are known a priori, the receiver can generate the corresponding noiseless pilot templates for each candidate distance $d\in\mathcal D$. The initial distance estimate is then obtained by least-squares matching between these templates and the received pilot blocks:
\begin{equation}
    \hat d^{(0)}
    =
    \arg\min_{d\in\mathcal D}
    \sum_{m=1}^{K_{\mathrm p}}
    \left\|
    \mathbf z_m-\boldsymbol{\mu}_m(d,\mathbf a^{\mathrm p})
    \right\|^2 .
    \label{eq:distance_ls_estimator}
\end{equation}
The superscript $(\cdot)^{(0)}$ indicates that this estimate initializes the subsequent iterative refinement. This estimator provides a low-complexity initialization for the distance-aware detector. Under equal-variance Gaussian block observations, it can also be interpreted as an approximate maximum-likelihood distance estimator.

\subsection{Distance-Aware Low-Complexity Symbol Detection}
\label{subsec:distance_aware_detection}

Given the initial distance estimate $\hat d^{(0)}$ from~\cref{eq:distance_ls_estimator}, the receiver constructs the corresponding distance-dependent block responses for data detection. For a symbol delay $\ell$, the sampled block response is defined as
\begin{equation}
\hat{\mathbf g}_{\ell}^{(0)}
\triangleq
\bigl[
g_{\ell N_s+qM-1}(\hat d^{(0)})
\bigr]_{q=1}^{X}
\in\mathbb{R}^{X},
\qquad \ell\in\mathbb{Z}_{\ge 0}.
\label{eq:estimated_block_cir}
\end{equation}
Here, $\ell=0$ corresponds to the current symbol contribution, whereas $\ell\ge 1$ represents post-cursor inter-symbol interference (ISI) from previously transmitted symbols.

Using the block observation model in~\cref{eq:block_obs_model}, the $m$-th observation block under the initial distance estimate can be written as
\begin{equation}
    \mathbf z_m
    =
    \boldsymbol{\mu}_m(\hat d^{(0)},\mathbf a)
    +
    \mathbf w_m .
    \label{eq:block_obs_detect_model}
\end{equation}
Combining the step-level response in~\cref{eq:step_mean_obs_model} with the block sampling rule in~\cref{eq:block_sample_index}, the corresponding noiseless block mean is
\begin{equation}
    \boldsymbol{\mu}_m(\hat d^{(0)},\mathbf a)
    =
    \sum_{\ell=0}^{m-1}
    N_{\mathrm{tx}} a_{m-\ell}
    \hat{\mathbf g}_{\ell}^{(0)} .
    \label{eq:block_mean_detect_model}
\end{equation}

To reduce detection complexity, each $X$-dimensional observation block is compressed into a scalar decision statistic using a combining vector $\mathbf c$:
\begin{equation}
    \tilde z_m
    =
    \mathbf c^\top \mathbf z_m,
    \qquad
    \mathbf c=\frac{1}{X}\mathbf 1_X ,
    \label{eq:block_compression}
\end{equation}
where $\mathbf 1_X\in\mathbb{R}^{X}$ denotes the all-one vector. Thus, $\mathbf c$ performs uniform averaging over the retained samples within one symbol interval. Applying the same combining vector to the block responses gives the compressed channel coefficient
\begin{equation}
    \tilde g_\ell^{(0)}
    =
    \mathbf c^\top \hat{\mathbf g}_{\ell}^{(0)},
    \qquad \ell\in\mathbb{Z}_{\ge 0}.
    \label{eq:compressed_cir}
\end{equation}

By applying the compression in~\cref{eq:block_compression} to the block observation model, the scalar observation becomes
\begin{equation}
    \tilde z_m
    =
    \tilde \mu_m(\hat d^{(0)},\mathbf a)
    +
    \tilde w_m,
    \label{eq:compressed_scalar_obs_model}
\end{equation}
where $\tilde w_m=\mathbf c^\top \mathbf w_m$ is the compressed noise term, and the compressed noiseless mean is
\begin{equation}
    \tilde \mu_m(\hat d^{(0)},\mathbf a)
    =
    \sum_{\ell=0}^{m-1}
    N_{\mathrm{tx}} a_{m-\ell}\tilde g_\ell^{(0)} .
    \label{eq:compressed_mean_detect_model}
\end{equation}

According to the scalar observation model in~\cref{eq:compressed_scalar_obs_model}, with the mean expanded in~\cref{eq:compressed_mean_detect_model}, the term with $\ell=0$ represents the current-symbol contribution, whereas the terms with $\ell\ge 1$ correspond to post-cursor ISI. We therefore employ a distance-aware decision-feedback equalization (DFE) with memory length $L$. During sequential data detection, the feedback symbol is defined as
\begin{equation}
\bar a_j^{(1)}
=
\begin{cases}
a_j^{\mathrm p}, & 1\le j\le K_{\mathrm p},\\
\hat a_j^{(1)}, & K_{\mathrm p}< j < m,\\
0, & j\le 0,
\end{cases}
\label{eq:feedback_symbol_def}
\end{equation}
where the superscript $(\cdot)^{(1)}$ in $\bar a_j^{(1)}$ and $\hat a_j^{(1)}$ indicates the first data-detection pass based on $\hat d^{(0)}$. For $m=K_{\mathrm p}+1,\ldots,K$, the ISI contribution from the previous $L$ symbols is estimated as
\begin{equation}
    \hat I_m^{(0)}
    =
    \sum_{\ell=1}^{L}
    N_{\mathrm{tx}} \tilde g_\ell^{(0)}
    \bar a_{m-\ell}^{(1)} .
    \label{eq:isi_estimate_dfe}
\end{equation}
where the summation starts from $\ell=1$ because $\ell=0$ corresponds to the current symbol. The data symbol is then detected after ISI cancellation as
\begin{equation}
    \hat a_m^{(1)}=
    \begin{cases}
    1, & \tilde z_m-\hat I_m^{(0)} \ge \eta^{(0)},\\
    0, & \text{otherwise},
    \end{cases}
    \label{eq:distance_aware_detector}
\end{equation}
with the midpoint threshold
\begin{equation}
    \eta^{(0)}
    =
    \frac{1}{2}N_{\mathrm{tx}}\tilde g_0^{(0)} .
    \label{eq:midpoint_threshold}
\end{equation}

\subsection{Iterative Joint Refinement Algorithm}
\label{subsec:iterative_joint}

The pilot-assisted distance estimate $\hat d^{(0)}$ enables the first distance-aware DFE-based data detection, yielding the initial data estimates $\{\hat a_m^{(1)}\}_{m=K_{\mathrm p}+1}^{K}$. Since the detected data symbols contain distance information through their observation blocks, they can be further exploited to refine the distance estimate. This motivates an iterative procedure that alternates between distance-aware data detection and data-aided distance estimation.

At iteration $t$, the current distance estimate $\hat d^{(t)}$ is used to recompute the distance-dependent block responses and compressed coefficients, following~\cref{eq:estimated_block_cir,eq:compressed_cir}. With these updated coefficients, the DFE rule in~\cref{eq:isi_estimate_dfe,eq:distance_aware_detector} produces the updated data estimates $\{\hat a_m^{(t+1)}\}_{m=K_{\mathrm p}+1}^{K}$. These estimates are combined with the known pilot symbols to form the reconstructed symbol sequence
\begin{equation}
\bar{\mathbf a}^{(t+1)}
=
[
a_1^{\mathrm p},\ldots,a_{K_{\mathrm p}}^{\mathrm p},
\hat a_{K_{\mathrm p}+1}^{(t+1)},\ldots,\hat a_K^{(t+1)}
]^\top .
\label{eq:reconstructed_symbol_sequence}
\end{equation}
Using the reconstructed sequence $\bar{\mathbf a}^{(t+1)}$, the receiver updates the distance estimate by matching all received blocks to their predicted noiseless templates:
\begin{equation}
    \hat d^{(t+1)}
    =
    \arg\min_{d\in\mathcal D}
    \sum_{m=1}^{K}
    \left\|
    \mathbf z_m-\boldsymbol{\mu}_m(d,\bar{\mathbf a}^{(t+1)})
    \right\|^2 .
    \label{eq:joint_distance_update}
\end{equation}
The updated distance estimate is then used for the next data-detection step, forming an alternating refinement between distance sensing and symbol detection. The process stops when the distance estimate no longer changes or when the maximum number of iterations is reached. The complete receiver procedure is summarized in~\cref{alg:isac_receiver}.

\begin{algorithm}[ht]
\caption{Pilot-Assisted Iterative Joint ISAC Receiver}
\label{alg:isac_receiver}
\begin{algorithmic}[1]
\STATE \textbf{Input:} $\{\mathbf z_m\}_{m=1}^{K}$, $\mathbf a^{\mathrm p}$, $\mathcal D$, $\mathbf c$, $L$, and $T_{\max}$
\STATE Initialize $\hat d^{(0)}$ using~\cref{eq:distance_ls_estimator}
\FOR{$t=0,1,\ldots,T_{\max}-1$}
    \STATE Apply the distance-aware DFE with $\hat d^{(t)}$ to obtain $\{\hat a_m^{(t+1)}\}_{m=K_{\mathrm p}+1}^{K}$
    \STATE Form $\bar{\mathbf a}^{(t+1)}$ by combining pilots and estimated data symbols
    \STATE Update $\hat d^{(t+1)}$ using~\cref{eq:joint_distance_update}
    \STATE \textbf{if} $\hat d^{(t+1)}=\hat d^{(t)}$ \textbf{then break}
\ENDFOR
\STATE \textbf{Output:} Final distance estimate $\hat d$ and detected data sequence $\hat{\mathbf a}$
\end{algorithmic}
\end{algorithm}

\begin{table}[ht]
    \centering
    \vspace{-0.3cm}
    \caption{Physical Channel Parameters.}
    \label{tab:channel_parameters}
    \renewcommand{\arraystretch}{1.2}
    \begin{tabular}{@{}lccc@{}}
        \toprule
            \textbf{Parameter} & \textbf{Symbol} & \textbf{Value}    & \textbf{Unit} \\ 
            \midrule
            Diffusion coefficient       & $D$ & \num{5e-11}    & \si{\metre\squared\per\second} \\ 
            Concentration of receptors  & $c_\text{p}$      & \num{1e-8}       & \si{M} \\
            Flow velocity               & $v$               & \num{10}         & \si{\micro\meter\per\second} \\
            Number of states            & $N$               & \num{301}      & -- \\
            Binding rate                & $k_\text{on}$     & \num{6e8}      & \si{ M^{-1}.s^{-1}} \\
            Unbinding rate              & $k_\text{off}$    & \num{3}        & \si{s^{-1}} \\
            Spatial step size           & $\Delta x$        & \num{1e-6}       & \si{\meter} \\
            Time step size              & $\Delta t$        & \num{8e-4}     & \si{s} \\
        \bottomrule
    \end{tabular}
    \vspace{-0.3cm}
\end{table}

\section{Numerical Results} 
\label{sec:results}

This section evaluates the proposed molecular ISAC framework through numerical simulations and examines both distance sensing and data detection performance.

\subsection{Simulation Setup}

The simulations are performed in MATLAB. Unless otherwise stated, the physical channel parameters and the communication/receiver parameters are given in~\cref{tab:channel_parameters} and~\cref{tab:comm_parameters}, respectively. Distance sensing is performed over the discrete candidate set $\mathcal D=\{130,135,140,145,150,155,160\}\,\si{\micro\meter}$. For each candidate distance $d\in\mathcal D$, the distance-dependent CIR in~\cref{eq:cir_def_model} is used to generate the block observation templates according to~\cref{eq:block_obs_model}. The receiver then performs pilot-assisted distance initialization, distance-aware DFE detection, and iterative joint refinement according to~\cref{alg:isac_receiver}. All reported results are averaged over $1000$ independent Monte Carlo trials unless otherwise stated.

\begin{table}[h]
    \centering
    \caption{Communication and Receiver Parameters.}
    \label{tab:comm_parameters}
    \renewcommand{\arraystretch}{1.2}
    \begin{tabular}{@{}lccc@{}}
        \toprule
        \textbf{Parameter} & \textbf{Symbol} & \textbf{Value} & \textbf{Unit} \\
        \midrule
        Symbol interval & $T_b$ & 12 & \si{\second} \\
        Sampling interval & $T_s$ & 2.4 & \si{\second} \\
        Pilot length & $K_{\mathrm p}$ & 2 & -- \\
        Data length & $K_{\mathrm d}$ & 1000 & -- \\
        Detector memory length & $L$ & 2 & -- \\
        Maximum iteration number & $T_{\max}$ & 5 & -- \\
        \bottomrule
    \end{tabular}
\end{table}

The distance sensing performance is measured by the sensing accuracy
\begin{equation}
    P_{\mathrm{acc}}
    \triangleq
    \Pr(\hat d=d_{\mathrm{true}}),
    \label{eq:sensing_accuracy}
\end{equation}
where $d_{\mathrm{true}}$ is the true TX--RX distance. The communication performance is measured by the bit error ratio (BER). To quantify the communication gain from iterative joint refinement, we define the relative BER reduction at iteration $t$ as
\begin{equation}
    \Delta_{\mathrm{BER}}^{(t)}
    \triangleq
    \frac{\mathrm{BER}^{(0)}-\mathrm{BER}^{(t)}}{\mathrm{BER}^{(0)}},
    \label{eq:relative_ber_reduction}
\end{equation}
where $\mathrm{BER}^{(0)}$ and $\mathrm{BER}^{(t)}$ denote the BER after pilot-based initialization and after the $t$-th refinement iteration, respectively. Thus, $\Delta_{\mathrm{BER}}^{(0)}=0$, and a larger value indicates a greater BER improvement over the initialization stage.

\subsection{Performance of Distance-Aware Sensing and Detection}

\begin{figure}[t]
    \centering
    \subfigure[\label{fig:accuracy} Distance accuracy]{%
        \centering
        \includegraphics[width=0.49\linewidth]{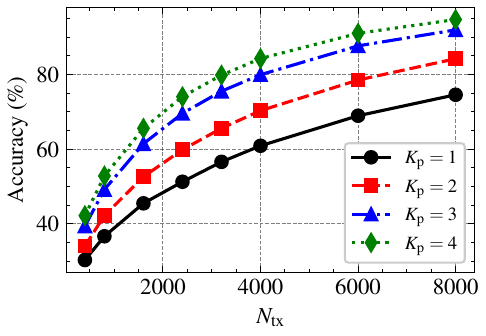}}
    \hfill
    \subfigure[\label{fig:ber_distance_awareness} BER performance]{%
        \centering
        \includegraphics[width=0.49\linewidth]{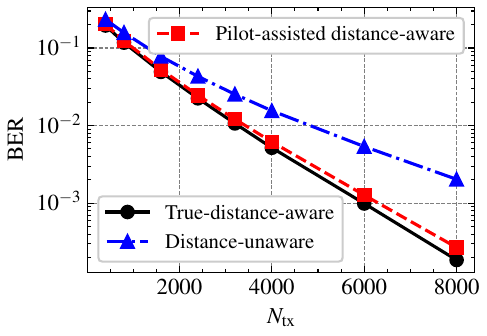}}
     \caption{Performance of distance-aware sensing and detection without iterative joint refinement. 
    (a) Distance sensing accuracy $P_{\mathrm{acc}}$ versus the number of released molecules $N_{\mathrm{tx}}$ for different pilot lengths $K_{\mathrm p}\in\{1,2,3,4\}$. 
    (b) BER versus $N_{\mathrm{tx}}$ for different detector designs, where the true-distance-aware DFE uses $d_{\mathrm{true}}=150\,\si{\micro\meter}$ and the distance-unaware DFE uses $d_{\mathrm{un}}=140\,\si{\micro\meter}$.}
    \label{fig:without_iteration}
    \vspace{-0.4cm}
\end{figure}

\Cref{fig:without_iteration} evaluates the sensing and detection performance before iterative refinement. As shown in~\cref{fig:accuracy}, the distance sensing accuracy increases with the number of released molecules $N_{\mathrm{tx}}$. This is because larger molecular releases improve the reliability of the received pilot blocks and make the distance-dependent templates more distinguishable. For a fixed $N_{\mathrm{tx}}$, increasing the pilot length $K_{\mathrm p}$ further improves sensing accuracy by providing more observations for the least-squares distance matching in~\cref{eq:distance_ls_estimator}. This gain is most visible in the low- and moderate-$N_{\mathrm{tx}}$ regimes.

\Cref{fig:ber_distance_awareness} compares three detector designs. The true-distance-aware DFE assumes perfect knowledge of the actual TX--RX distance, $d_{\mathrm{true}}=150\,\si{\micro\meter}$, and serves as a benchmark. The proposed pilot-assisted distance-aware DFE uses the initial distance estimate $\hat d^{(0)}$ obtained from~\cref{eq:distance_ls_estimator}. The distance-unaware DFE uses a fixed mismatched distance $d_{\mathrm{un}}=140\,\si{\micro\meter}$. The results show that distance awareness improves data detection. The proposed pilot-assisted distance-aware DFE approaches the true-distance-aware benchmark as $N_{\mathrm{tx}}$ increases, whereas the distance-unaware DFE suffers from channel mismatch and yields a higher BER. These results confirm that distance sensing provides useful channel information for communication in the proposed molecular ISAC receiver.

\subsection{Impact of Iterative Joint Refinement}

\begin{figure}[t]
    \centering
    \subfigure[\label{fig:distance_accuracy_vs_iteration:a} Distance accuracy]{%
        \centering
        \includegraphics[width=0.49\linewidth]{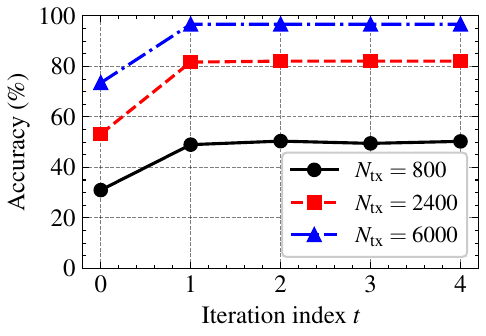}}
    \hfill
    \subfigure[\label{fig:relative_ber_reduction_vs_iteration} Relative BER reduction]{%
        \centering
        \includegraphics[width=0.49\linewidth]{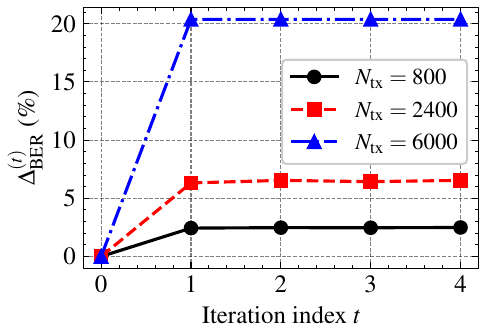}}
    \caption{Impact of iterative joint refinement on sensing and communication performance for $N_{\mathrm{tx}}\in\{800,2400,6000\}$.  
    (a) Distance sensing accuracy $P_{\mathrm{acc}}$ versus iteration index.  
    (b) Relative BER reduction versus iteration index, where the reduction is normalized by the initialization-stage BER of each curve.}
    \label{fig:iteration}
    \vspace{-0.4cm}
\end{figure}

\Cref{fig:iteration} evaluates the impact of iterative joint refinement on both sensing and BER performance. As shown in~\cref{fig:distance_accuracy_vs_iteration:a}, the distance sensing accuracy improves after refinement, with the largest gain achieved in the first iteration. The improvement becomes more pronounced as $N_{\mathrm{tx}}$ increases, since stronger molecular observations lead to more reliable data decisions and more accurate data-aided distance re-estimation.

The corresponding BER improvement is shown in~\cref{fig:relative_ber_reduction_vs_iteration}. The relative BER reduction increases rapidly after the first refinement iteration, especially for larger $N_{\mathrm{tx}}$. This confirms the closed-loop benefit of the proposed receiver: improved distance estimates provide more accurate channel templates for DFE detection, while improved data decisions support more reliable distance refinement. For smaller $N_{\mathrm{tx}}$, the gain is limited by less reliable sensing and detection. The curves in both subfigures quickly saturate, indicating that the proposed alternating refinement converges rapidly and captures most of the available gains within the first few iterations.

\section{Conclusion} \label{sec:conclusion}

This paper presented a first step toward a broader molecular ISAC framework in which common molecular observations are exploited for both physical-parameter sensing and data detection. As a representative instantiation, we considered a microfluidic MC channel and focused on TX--RX distance sensing. By embedding the TX--RX distance into a distance-parameterized Markov state-space model, we obtained distance-dependent CIRs and a symbol-wise block observation model for OOK signaling. Based on this model, we developed a pilot-assisted low-complexity receiver that combines distance initialization, distance-aware DFE detection, and iterative joint refinement. Numerical results showed that the proposed receiver achieves accurate distance sensing and improved BER performance, and that iterative refinement enables sensing and communication to mutually benefit from each other. These results demonstrate the potential of microfluidic MC as a representative platform for molecular ISAC. Future work will extend the framework to continuous-valued parameter estimation beyond the discrete candidate set. The same principle can also be generalized to broader molecular ISAC scenarios involving flow velocity, reaction kinetics, receiver-state variations, higher-dimensional channel geometries, and multi-parameter sensing.

\bibliographystyle{IEEEtran}
\bibliography{IEEEabrv,refs}
\end{document}